\documentstyle[12pt,openbib]{article}

\begin{document}

\title{\bf Absence of Cooper-type bound states in three- and 
few-electron systems\thanks{To appear in European Physical Journal B}}

\author{Sadhan K. Adhikari$^1$\thanks{John Simon Guggenheim Memorial 
Foundation Fellow}
  and T. Frederico$^2$\\
$^1$Instituto de F\'{\i}sica Te\'orica, Universidade Estadual 
Paulista,\\ 01.405-900 S\~ao Paulo, S\~ao Paulo, Brazil\\
$^2$Departamento de F\'{\i}sica, Instituto Tecnol\'ogico Aeronautica, \\
Centro T\'ecnico Aeroespacial, \\ 12228-900 S\~ao Jos\'e dos Campos, 
S\~ao Paulo, Brazil}


\maketitle

\begin{abstract}

It is shown that the appearance of a fixed-point singularity in the kernel of
the two-electron Cooper problem is responsible for the formation of the
Cooper pair for an arbitrarily weak  attractive interaction between two
electrons. This singularity is absent in the problem of 
 three and  few superconducting electrons at zero temperature 
 on the full  Fermi sea. Consequently, such three- and few-electron
systems on the full  Fermi sea do not form Cooper-type bound states for an
arbitrarily weak  attractive pair interaction.

PACS Numbers {74.20.Fg}


\end{abstract} 

\newpage

For an arbitrarily weak residual attractive interaction, at zero 
temperature two electrons over
the full Fermi sea spontaneously form a bound Cooper pair \cite{e}.  These
Cooper pairs lay the foundation of  the microscopic Bardeen-Cooper-Schreiffer
(BCS) theory of superconductivity \cite{8}. Flux quantization and other
experimental evidences support the formation of Cooper pairs and their role
in the BCS theory.   Appearance of a
fixed-point  singularity in the kernel of the momentum space Schr\"odinger
equation is responsible for the formation of Cooper pairs for arbitrarily
weak attractive interaction. Similar singularity also appears in a
general
two-fermion problem in vacuum in one and two dimensions and is responsible,
in these cases, for the formation of two-fermion bound states for very weak
attractive interactions. This singularity  is absent in the two-fermion
problem in vacuum in three dimensions, and hence in that case there is no
two-fermion bound state for a very weak attractive interaction.

The possibility of few-fermion  clustering at low temperature 
  is of recent interest \cite{4}
in the BCS to Bose crossover problem \cite{NSR}  in the Fermi liquid 
model of  not only a free electron gas 
but several other fermionic systems such as  nuclear matter, 
neutron matter (neutron stars), and electron-hole
systems. Although Cooper pairing is supposed to dominate the weak-coupling 
BCS limit, it is not clear that 
few- and multi-electron  clustering are not allowed in different spin and 
angular momentum states for coupling simulating a crossover from the BCS to 
Bose problem.  This possibility  is of  concern as in the
low-density region of nuclear matter it has been shown \cite{4} that at low 
temperatures the dominant part of nuclear matter will form $\alpha$
particles which are more strongly bound than deuterons.
Hence,  spin-triplet (deuteron) Cooper pairing and subsequent condensation of
deuterons 
has to compete with $\alpha$-particle Cooper quartetting and subsequent 
condensation. 
Here, in view of the  study  of Cooper-type quartetting in nuclear
matter, 
in this work we consider the possibility of few-electron clustering 
within the Fermi liquid model of electrons. The principal difference
between the nuclear matter and the superconducting 
electron gas problem is that in the former the interaction is much stronger
than the weak residual attraction in case of electrons. 
So far existence of few-electron Cooper-type bound states have not been 
experimentally confirmed. However, 
 if few-electron Cooper-type 
  bound states are allowed in a specific spin/angular 
momentum state (for example, in an exotic high-$T_c$ superconductor), 
they should be correctly accomodated in any microscopic theory of
superconductivity.

At zero temperature, a Cooper-type consideration 
of the three-electron problem on the full  Fermi sea
for a very weak attractive  interaction shows that a  fixed-point
singularity of the type discussed above 
 is absent in the kernel of the three-particle equation in
momentum space. Hence there is no Cooper-type bound state for the
three-electron system in any space dimension. Similar results should hold for
clustering of $n$ electrons with $n > 3$ on the full Fermi sea.

We  consider the three-electron problem on the full  Fermi sea with an
arbitrarily weak  two-electron interaction in $S$ wave.  In this case, for
the state with total orbital angular momentum  of the three-electron system
$L= 0$, obviously the effective interaction is repulsive because of the Pauli
exclusion principle and there is no bound state. This is because a third
electron, with spin up or down, can not approximate a singlet electron pair
with one spin-up and one spin-down electron and form a three-electron bound
state for $L=0$, as the antisymmetrization of the three-electron wave
function has to follow entirely from its spin part. However, for the
three-electron state with total orbital angular momentum $L$ odd, the
antisymmetrization of the wave function follows partly from its spin part and
partly from its orbital part. In this case,
 there could be an effective attraction
between the three  electrons in the state with $L=1$, which could form
a bound state. We write the Faddeev
equation \cite{f} 
for the three-electron problem on the full  Fermi sea and find that
the effective three-electron interaction is attractive  for odd $L$ states
and repulsive for even $L$ states for a very weak two-electron attractive
interaction in $S$ state.  However, the Cooper-type singularity does not
appear in the kernel  of the Faddeev equation. Hence there are no Cooper-type
three-electron bound states for very weak attractive interactions.

In Sec. II we present a discussion of the two-electron problem (a) in vacuum
and (b) on the full Fermi sea for an arbitrarily weak  attractive potential.
We show that the presence of a fixed-point singularity  in the kernel of the
problem in momentum space is responsible for the formation of a bound state
(a) in vacuum in one and two dimensions and (b) on the  full Fermi sea in any
dimension.  In Sec. III we formulate the three-electron problem on the full
Fermi sea for an arbitrarily weak  attractive pair potential. We find that a
fixed-point  singularity does not appear in the kernel of the momentum-space
Faddeev equation in this case and there is no three-electron bound state for
an arbitrarily weak attractive pair potential.

\section{The Two-Electron Problem}

For two electrons, each of mass $m$, in the center of mass frame the single
(two) particle energy is given by $\epsilon_q = q^2$ ($2q^2$), in units
$\hbar=2m=1$, where $q$ is the wave number.  We consider a purely attractive
weak residual
$S$-wave short-range  separable potential between electrons:
\begin{equation}
V(p, q)=-\lambda  g(p) g(q) .\label{0}
\end{equation}
Because of the presence of the lattice, the superconducting electrons
experience a pairwise weak finite-range residual attraction which is modelled
by the above potential. In the conventional BCS model the 
potential form factors $g(p)$ 
are 
set equal to unity and the range of the potential is introduced by the 
Debye cut off in momentum space \cite{8}. The analysis and conclusion 
of this work are  independent of
this specific form of the potential employed.   
For Cooper pairing in zero (even) orbital angular momentum state(s), the  
allowed spin state of the two-electron system is $S = 0$ by
Pauli exclusion principle. Hence this potential acts in the $^1S$ state: the
$S$-wave spin-singlet state.  The two-electron problem, with this potential
at energy $E$, is given by the following equation
\begin{equation}
f(q) = \int_c^\infty d^{\cal D}p \frac{V(q,p)f(p)}{E-2p^2}, \label{1}
\end{equation}
where $c=0$ in vacuum, and $c=k_F$, the Fermi momentum, for the Cooper
problem. Here $G\equiv (E-2p^2)^{-1}$ is the Green function, $f(q)$ the 
bound-state form
factor, $E$ the two-electron energy and ${\cal D}$ is the dimension of the
space.  For a two-electron bound state in vacuum, we have $E\le 0$  and for
the bound state over the full Fermi sea, the condition $E\le 2k_F^2$ is to be
satisfied.

In vacuum, for $\lambda  \to 0$, the Green function develops a singularity at
the lower limit as the binding energy $\alpha^2 (\equiv -E)\to 0.$ In this
limit, for the above separable potential,  Eq. (\ref{1}) reduces to
\begin{equation} 1 = \lambda C\int_0^\infty p^{{\cal D}-1}dp
\frac{g^2(p)}{2p^2}, \label{2}
\end{equation}
where $C=4\pi$ ($2\pi$, 2)  for ${\cal D} = 3$ (2, 1).  In three space
dimensions, (${\cal D}=3$), for usual well-behaved
potential form factors $g(p)$, the
integral in Eq. (\ref{2}) is finite and it is impossible to satisfy condition
(\ref{2}) in the limit $\lambda  \to 0$. Hence there are no bound states in
vacuum for very weak potentials in three dimensions. However, in one and two
dimensions, the integral in Eq. (\ref{2}) is infinite and one can satisfy
condition (\ref{2}) in the limit $\lambda  \to 0$ and one can have bound
states in one and two dimensions for arbitrarily weak attractive potentials.

On the  full Fermi sea,  as $\lambda  \to 0$, the Green function also
develops a singularity at the lower limit as the Cooper pair binding energy
$\alpha^2  (\equiv 2k_F ^2 -E) \to 0$.  In this limit,   for the above
separable potential,  Eq. (\ref{1}) reduces to \begin{equation} 1 = \lambda C
\int_{k_F}^\infty p^{{\cal D}-1}dp
\frac{g^2(p)}{2(p^2-k_F^2)}. \label{3}
\end{equation}
Equation (\ref{3}) represents the condition for Cooper instability.  The
integral in Eq. (\ref{3}) is infinite in any space dimension for usual 
well-behaved form factors $g(p)$ and Eq.
(\ref{3}) can be satisfied in the limit $\lambda \to 0$ for any $k_F \ne 0$.
For $k_F=0$, the singularity in the Green function
of Eq. (\ref{3}) at the lower limit of the integral is cancelled by a zero in
the phase space $p^{{\cal D}-1}dp$ for ${\cal D} =3.$ This singularity for
$k_F=0$ survives in the case of ${\cal D} = 1$ and 2 and is responsible for
the two-electron bound state in vacuum for any arbitrarily weak attractive
potential.

We base our discussion on equations of type 
(\ref{3}) which led Cooper to his 
conclusion on pair formation. A more complete discussion could 
be based on a linearized version of the gap equation.
The present    simplification does not 
change anything in our qualitative discussion on the existence  of 
 Cooper-type states based of the appearance of singularities. 
The essential difference between the two approaches is in the 
use of the right phase space factors, which is not expected to change 
the general criteria for the appearance of singularities as considered
here for two- and few-electron systems.

Consequently,  one can have Cooper pairs in any  space dimension for
electrons interacting via a weak attractive  interaction over the full Fermi
sea. We have used a separable potential to reach the above conclusion.  As
the arguments are based on the existence of a singularity in the kernel,  the
conclusion should hold  for any short-range potential.  For potential
(\ref{0}), the two-electron $t$ matrix for dynamics on the full Fermi sea has
the following analytic form
\begin{equation}
 t(p,q,E) = g(p) \tau(E) g(q),\label{5}
\end{equation} 
where 
\begin{equation}
\tau(E) =  \biggr[-\frac{1}{\lambda} - \int_{k_F} ^\infty
d^{\cal D}q  g^2(q) (E-2q^2)^{-1}
\biggr]^{-1}.  \label{6}
\end{equation}
The use of  separable potential (\ref{0}) facilitates the solution of the
three-electron problem on the full Fermi sea, which we take up in the next
section.

\section{The Three-Electron Problem}

The simplest of the three-electron problem on the full Fermi sea,
 that we consider here in some
details, is the one where they interact via $S$-wave pair potential
(\ref{0}) in the $^1S$ state.  This is the problem of three superconducting
electrons on the top of the full Fermi sea at zero temperature and 
 is in effect a many-body problem 
involving all the electrons and the lattice. However, the many-body nature of
the problem introduces only minor changes over the three-electron problem 
in vacuum.
As in the two-electron Cooper problem, one has a weak attractive interaction 
between the electrons in place of the Coulomb repulsion in vacuum
with an appropriate truncation of the phase space in 
momentum space consistent with the Pauli principle. So effectively one has 
the problem of three  electrons under the action of a weak attractive 
interaction in vacuum subject to the above-mentioned
restrictions on the momentum-space phase space arising from  the many-body  
nature of the problem.

The most likely assignment of the quantum number
$(LSJ)$ (total orbital angular momentum, total spin, and total angular
momentum) for the bound state of the three electrons on the full Fermi sea
in this case is $(L=1, S=1/2, J=1/2$ and 3/2).
The three-electron bound state can be visualized as the bound state of the
third (spectator) particle with the spin-singlet bound state of the first two
particles (1 and 2). For the spectator particle to be bound to the singlet
pair, lowest value for its angular momentum state relative to the pair should
be ${\cal L} =1$.  The value ${\cal L} =0$ 
is not allowed by the Pauli exclusion
principle.  This is why the lowest probable value of $L$ is 1. The only
possible value for $S$ is 1/2, so that there are two degenerate $J$ values
1/2 and 3/2.  In the following we shall consider only this state. If, in
addition, one allows a two-electron potential in the 
$^3P$ state, one can have a
three-electron bound state for $(LSJ)$ = (1,3/2,1/2), (1,3/2,3/2), (0,
1/2,1/2)  etc.  A complete analysis of these states, in the context of the
three-neutron system,  has been given by Mitra \cite{m}.
   
The $t$ matrix of Eq. (\ref{5}) acts as an effective potential in the
three-electron Faddeev equation  in three dimensions, which in this case can
be written as \cite{f,a}
\begin{equation}
F({\bf q},E) =  2\tau(E-3q^2/2)\int  d^3 p K({\bf q, p})F({\bf p},E),
\label{7}\end{equation} where
\begin{equation}K({\bf q, p})=\chi \frac {g(|{\bf p +q/2}|)g(|{\bf
q+p/2}|)}{E-A},
\label{70}
\end{equation}
with $A=p^2+q^2+({\bf p + q})^2$. Now $E$ is the three-electron energy and
for the bound state on the full Fermi sea $E\le 3k_F^2.$ Here, for $L=1$ and
$S=1/2$, the spin  coupling coefficient $\chi= -1/2$, and $p, q, |{\bf p+ \bf
q}| > k_F$. We should now stick to a specific total angular momentum state
${\cal L}$ (of the spectator particle) and take a partial-wave projection of
Eq. (\ref{7}) in ${\cal L}$ before solving it. The partial-wave projection of
Eq. (\ref{7})  is given by
\begin{equation}
F_{\cal L}({ q},E)=4\pi \tau(E-3q^2/2)\int_{k_F}^\infty p^2 dp K_{\cal L}(q,p)
F_{\cal L}({ p},E),
\label{8}
\end{equation}
where
\begin{equation}
K_{\cal L}(q,p)=\int_{-1}^{1}dx{P_{\cal L}(x)}K({\bf q, p}){\Theta(A-3k_F^2)
\Theta(q^2-k_F^2)}\label{9}
\end{equation}
is the partial-wave kernel for this problem. Here $x$ is the cosine of the
angle between ${\bf p}$ and ${\bf q}$, $P_{\cal L}(x)$ is the Legendre
polynomial, and $\Theta(x)=1$ for $x\ge 0$ and = 0, otherwise.  Equation
(\ref{9}) with $k_F=0$ is the usual kernel of the Faddeev equation.  A direct
calculation has revealed that, for an attractive potential, the kernel
(\ref{9}) is positive (negative) definite for ${\cal L}$ odd (even). Hence
the three-body equation (\ref{8}) is purely attractive for odd ${\cal L}$ and
we shall consider ${\cal L}= 1$ below.

For a detailed study of this problem, one should consider the partial-wave
Eq. (\ref{8}). However, the full equation (\ref{7}) reveals  the essential
interesting feature. The Green function of Eq. (\ref{70}), unlike in the
two-electron problem, does not have any fixed-point singularity for all $q$.
In the arbitrarily weak attractive potential limit $\lambda \to 0$, the
function $\tau $ of Eqs. (\ref{5}) and (\ref{7}) also tends to zero. Then one
faces the question whether Eq. (\ref{7}) permits a nontrivial solution in
this limit, which could correspond to a weakly bound three-electron state. As
the binding energy of the three-electron system is
expected to tend to zero,  the
appropriate $E$ in this equation is $E=3k_F^2$. One can see from Eq.
(\ref{7}) that the energy denominator in this equation does not have a
fixed-point singularity, as in the two-electron case.   Hence the integral in
Eq. (\ref{7}) is finite and its right hand side  is always zero in the
weak-coupling limit as $\tau \to 0.$ This means that the form factor $F(q,E)$
is identically zero in the weak-coupling limit and there is no three-electron
bound state.

The situation does not change after the partial-wave projection. The kernel
$K_{\cal L}(q,p)$ of Eq. (\ref{9}) for ${\cal L} =1$ develops weak and
integrable moving logarithmic singularities at $2p^2+2q^2\pm 2pq = 3k_F^2$,
as opposed to fixed-point singularities in the two-electron problem. As these
singularities are integrable, the integral on the right hand side of Eq.
(\ref{8}) is always finite, so that, in the limit of weak interaction, as
$\tau \to 0$, this equation has the only trivial solution $F(q,E)=0$. Hence
there are no Cooper-type 
three-electron bound state in this limit.  However,  the kernel
is attractive in this case and one can have a three-electron bound state as
the strength of the potential is increased.  

The present general discussion is based on the existence of a fixed-point
singularity and not on some specific property of the system. Hence it should be
applicable to other similar problems. For example, identical behavior is
expected in one and two dimensions and other ($LSJ$) three-electron bound
states on the full Fermi sea. So  there are no Cooper type bound states in
any space dimension for the three-electron system for an arbitrarily weak
two-electron potential.

The present discussion can easily be extended to clustering of $n$ electrons
($n\ge 4$) on the full Fermi sea. For $n=4$, the problem 
 should be treated by a four-body dynamical equation.
The kernel of such equation will
involve four- and three-electron Green functions, none of which could have a
fixed-point singularity. Hence, by similar arguments, the four-electron
Cooper-type bound states are not also allowed for an arbitrarily weak
two-electron potential.

\section{Summary}

We have shown that the appearance of a fixed-point singularity in the kernel
of the momentum-space Schr\"odinger equation is responsible
for the existence of a bound state for an arbitrarily weak attractive
potential in the zero-temperature limit.  
The formation of a Cooper pair in any space dimension and of a
two-electron bound state in vacuum in one and two space dimensions under the
action of a very weak attractive 
potential is due to  the appearance of the above
fixed-point singularity. Similar fixed-point singularities are not expected
to appear in the  $n$-electron $ (n > 2)$ problem 
and there are no
Cooper-type bound states in these cases for arbitrarily weak attractive
potentials in the zero-temperature limit. 
However, with stronger attractive potentials three and
few-electon  clusters can be formed on the full Fermi sea as has been
discussed recently in the low-density limit of nuclear matter at low
temperatures \cite{4}.

We thank Conselho Nacional de
Desenvolvimento Cient\'{\i}fico e Tecnol\'ogico and Funda\c c\~ao de Amparo
\`a Pesquisa do Estado de S\~ao Paulo for financial support.


\begin{thebibliography}{99}

\bibitem{e} L. N. Cooper, Phys. Rev. {\bf 104}, 1189 (1956).

\bibitem{8} J. Bardeen, L. N. Cooper, and J. R. Schrieffer, Phys. Rev. 
{\bf 108}, 1175 (1957); J. R. Schrieffer, {\em Theory of Superconductivity},
(Benjamin, New York, 1964).

\bibitem{NSR}
P. Nozi\`eres and S. Schmitt-Rink,
J. Low Temp. Phys. {\bf 59}, 159 (1985);
S. K. Adhikari and A. Ghosh, Phys. Rev.B {\bf 55}, 1110 (1997);
J. Phys.: Cond. Mat. {\bf 10}, 135 (1998); A. Ghosh and S. K. Adhikari,
Euro. Phys. J. B {\bf 2}, 31 (1998).

\bibitem{4}G.\ R\"opke, A.\ Schnell, P.\ Schuck, and P.\ Nozi\`{e}res,
Phys. Rev. Lett. {\bf 80}, 3177 (1998). 




\bibitem{f}L. D. Faddeev, {\it Mathematical Aspects of the Three-Body 
Problem in the Quantum Scattering Theory}, (Israel Program of Scientific 
Translation, Jerusalem, 1965).

\bibitem{m} A. N. Mitra and V. S. Bhasin, Phys. Rev. Lett. {\bf 16},
523 (1966); A. N. Mitra, Phys. Rev. {\bf 150}, 839 (1966).


\bibitem{a} S. K. Adhikari and K. L. Kowalski, {\it Dynamical Collision 
Theory and its Applications}, (Academic Press, Boston, 1991).




\end{thebibliography}
\end{document}